\ifx\mnmacrosloaded\undefined \input mn \input epsf\fi

\newif\ifAMStwofonts

\ifCUPmtplainloaded \else
  \NewTextAlphabet{textbfit} {cmbxti10} {}
  \NewTextAlphabet{textbfss} {cmssbx10} {}
  \NewMathAlphabet{mathbfit} {cmbxti10} {} 
  \NewMathAlphabet{mathbfss} {cmssbx10} {} 
  \ifAMStwofonts
    \NewSymbolFont{upmath} {eurm10}
    \NewSymbolFont{AMSa} {msam10}
    \NewMathSymbol{\upi}     {0}{upmath}{19}
    \NewMathSymbol{\umu}     {0}{upmath}{16}
    \NewMathSymbol{\upartial}{0}{upmath}{40}
    \NewMathSymbol{\leqslant}{3}{AMSa}{36}
    \NewMathSymbol{\geqslant}{3}{AMSa}{3E}

    \let\leq=\leqslant 
     \let\ge=\geqslant
  \else
    \def\umu{\mu}
    \def\upi{\pi}
    \def\upartial{\partial}
  \fi
\fi

\pageoffset{-2.5pc}{0pc}

\loadboldmathnames



\pagerange{0--0}    
\pubyear{0000}
\volume{000}

\begintopmatter  

\title{Evolutionary calculations of carbon dredge-up in helium envelope
       white dwarfs}
\author{James MacDonald$^{1}$,
        Margarita Hernanz$^{2}$ and
        Jordi Jos\'e$^{3}$}
\affiliation{$^{1}$ Department of Physics and Astronomy, University of 
                    Delaware, Newark, DE 19716}
\affiliation{$^{2}$ Institut d'Estudis Espacials de Catalunya, 
                   CSIC Research Unit, Edifici Nexus-104, 
                   C/Gran Capit\`a, 2-4, E-08034 Barcelona, SPAIN}
\affiliation{$^{3}$ Departament de F\'\i sica i Enginyeria Nuclear (UPC), 
                   Avda. V\'\i ctor Balaguer, s/n, 
                   E-08800 Vilanova i la Geltr\'u, SPAIN}
\shortauthor{J. MacDonald, M. Hernanz and J. Jos\'e}
\shorttitle{Carbon dredge-up in helium envelope white dwarfs}


\acceptedline{Accepted ???? . Received 1997 ; in original form 1997}

\abstract {We investigate the evolution of cooling helium atmosphere white
dwarfs using a full evolutionary code, specifically developed for following
the effects of element diffusion and gravitational settling on white dwarf
cooling. The major difference between this work and previous work is that
we use more recent opacity data from the OPAL project. Since, in general,
these opacities are higher than those available ten years ago, at a given
effective temperature, convection zones go deeper than in models with older
opacity data. Thus convective dredge-up of observationally detectable
carbon in helium atmosphere white dwarfs can occur for thicker helium layers
than found by Pelletier et al (1986). We find that the range of observed C
to He ratios in different DQ white dwarfs of similar effective temperature
is well explained by a range of initial helium layer mass between $10^{-3}$
and $10^{-2} M_{\odot}$, in good agreement with stellar evolution theory,
assuming a typical white dwarf mass of $0.6 M_{\odot}$.
We also predict that oxygen will be present in DQ white dwarf atmospheres 
in detectable amounts if the helium layer mass is near the lower limit
compatible with stellar evolution theory. Determination of the oxygen
abundance has the potential of providing information on the profile of
oxygen in the core and hence on the important
$^{12}$C$(\alpha,\gamma)^{16}$O reaction rate.}


\maketitle  

\section{Introduction}

DQ white dwarfs have He-rich atmospheres, with the presence of a trace of 
carbon. The results of model atmosphere calculations for 24 DQ white dwarfs 
have been compiled by Weidemann \& Koester (1995). Effective temperatures, 
$T_{\rm eff}$, range from 12,500 K to 6,700 K. Log ($n$(C)/$n$(He)) 
values determined from visual spectra range from $-6.2$ to $-1.5$. Analysis 
of 
IUE spectra have pushed the lower limit to $-7.3$. The generally accepted 
explanation for the presence of trace carbon in these cool He white dwarf 
atmospheres is that carbon from the core diffuses outwards 
to where convection dredges it to the surface. The first detailed 
calculations 
of this process were presented by Pelletier et~al. (1986) (hereafter P86), 
who show how, in the hotter phases, carbon diffuses upwards from the core 
to be met eventually by the base of the surface convection zone 
that develops due to the increase in opacity following recombination of 
helium as the white dwarf cools. With further cooling more carbon diffuses 
upwards and the base of convection zone moves deeper into the star, further 
enriching the outer layers with carbon. A key finding of this research 
is that the maximum depth of the base of the convection zone nearly 
coincides with a change in ionization state of the carbon below the 
convection 
zone. Since the slope of the carbon profile depends on the relative 
ionization 
states of carbon and helium, this reduction in ionization state of 
carbon results in a steepening profile and a reduction in the carbon 
abundance 
in the convection zone. It is this change in ionization structure that 
causes 
a reduction in observed $n$(C)/$n$(He) values at lower temperatures, and the 
non-detection of carbon below 6,000 K. Since the thickness 
of the helium layer regulates the effective temperature at which the base of 
the convection zone reaches regions of diffusively enhanced carbon, the 
observations of DQ white dwarfs place limits on the amount of helium in the 
white dwarf, that can be compared to the predictions of stellar 
evolution theory. P86 concluded that models for a white dwarf mass of 
$M_{*} = 0.6 M_{\odot}$ and a helium envelope mass, $M_{\rm env}$, in the 
range $10^{-4}$ to $10^{-3.5}$ $M_{*}$ gave the best fit to the contemporary 
observational data. This is in conflict with evolutionary models for post 
AGB stars that predict that $M_{\rm env}$ should be in the range $10^{-3}$ 
to $10^{-2} M_{\odot}$ for this core mass. To complicate the picture 
further, models of the non-radial g-mode pulsations of the DBV star GD~358 
fit the observations best for $M_{*} = 0.61 M_{\odot}$ and 
$M_{\rm env} = 10^{-5.7 \pm 0.3} M_{*}$ (Winget et~al. 1994, 
Bradley \& Winget 1994), which is significantly less than both the estimates 
from the DQ white dwarf models of P86 and post AGB evolution models.  

In this paper, we have reinvestigated the evolution of cooling helium 
atmosphere white dwarfs for a range of $M_{*}$ and $M_{\rm env}$, using a 
full evolutionary code, specifically developed for following the 
effects of element diffusion and gravitational settling on white dwarf 
cooling. The major difference between this work and that of P86 is that here 
we use the more recent opacity data from the OPAL project (Iglesias \& 
Rogers 
1993). Since, in general, these opacities are higher than those available 
ten 
years ago, at a given effective temperature, convection zones go deeper 
than in models with older opacity data. Thus convective dredge-up of 
observationally detectable carbon in helium atmosphere white dwarfs can 
occur 
for thicker helium layers than found by P86. We find that the range of 
observed 
C to He ratios in different DQ white dwarfs of similar effective temperature 
is well explained by a range of $M_{\rm env}$ between $10^{-3}$ and 
$10^{-2} M_{\odot}$, in good agreement with stellar evolution theory, 
assuming a typical white dwarf mass of $0.6 M_{\odot}$.

The details of the evolutionary code are given in the next section. In 
section 3, we present our results for the evolution of the abundance 
profiles 
and compare them with observations and current ideas for the evolution of 
possible white dwarf progenitors. Our conclusions are given in section 4.

\section{Modeling diffusion and convective mixing in cooling white dwarfs}

To study the evolution of the carbon distribution in helium atmosphere white 
dwarfs, we have added routines that solve the equations describing element 
diffusion and gravitational settling to a stellar evolution code 
specifically 
designed for following the cooling history of white dwarfs. We restrict 
attention to mixtures of helium, carbon and oxygen. The existence of OPAL 
opacities for arbitrary mixtures of these three elements allows us to 
calculate consistently the effects of element diffusion on the optical 
thickness of the non-degenerate envelope, which controls the rate of cooling 
of the white dwarf. These effects were not included by P86. The initial 
model consists of a core of uniform composition, surrounded by an envelope 
also of uniform composition. The initial temperature profile is determined 
by computing a thermal equilibrium model, in which radiative and neutrino 
losses are balanced by an artificial constant energy rate, 
$\varepsilon_{c}$. 
We study three sets of models, labeled HE, LO, and PG. To identify our 
sequences, we use an alphanumeric label that consists of three parts. The 
first part identifies the core mass, the second part the particular set of 
models to which the sequence belongs, and the third part identifies the 
envelope mass. For example, the P6HE6 sequence is part of the HE set of 
models and has a core mass of $0.6 M_{\odot}$ and envelope mass of 
approximately $10^{-6} M_{\odot}$. The HE set is the largest set of models 
and explores how the degree of carbon and oxygen dredge-up depends on core 
mass and envelope mass. The LO set has a lower core oxygen abundance than 
the 
HE set and is used to explore the relation between the oxygen abundance in 
DQ 
photospheres and the oxygen abundance in the core. The PG sequence has 
envelope composition characteristic of the PG1159 stars and is used to test 
the idea that some DQ white dwarfs are descendants of the PG1159 stars. 
In particular, Dehner \& Kawaler (1995) have suggested that the unusual 
DBQv GD 358 is such an object. The properties of our initial models are 
given in Table~1. $M_{\rm core}$, $X_{\rm C,core}$, $X_{\rm O,core}$ and 
$M_{\rm env}$, $X_{\rm C,env}$, $X_{\rm O,env}$ are the initial 
masses, carbon mass fraction and oxygen mass fraction of the CO core and 
He-rich envelope respectively. Our range of $M_{\rm env}$ values at 
$0.6 M_{\odot}$ was chosen to include those from the asteroseismological 
studies of GD 358 up to the largest values consistent with stellar evolution 
models for the white dwarf progenitor.

We stop our evolutionary calculations when the temperature at zero 
optical depth is 6,000 K, which is the lowest temperature in the OPAL 
opacity tables. This corresponds to $T_{\rm eff} = 7,000$ K.

The treatment of diffusion is fairly straightforward. We use an approach 
that is applicable to a general mixture of an arbitrary number of elements. 
We neglect thermal diffusion, which has been shown to be negligible for 
white dwarfs by Iben \& MacDonald (1985), and Paquette et~al. (1986b) and 
radiative levitation. Although radiative levitation is important for 
determining the photospheric composition of hot white dwarfs (Fontaine \& 
Michaud 1979, Vauclair, Vauclair \& Greenstein 1979), it is entirely 
negligible in the interior and also in the photospheric regions at the 
effective temperatures of the DQ white dwarfs. For simplicity, we also 
neglect magnetic fields and stellar rotation. The composition profiles are 
then determined by the competition between gravity, partial pressure 
gradients 
and induced electric fields. The diffusion velocities satisfy (Curtiss \& 
Hirschfelder 1949, Aller \& Chapman 1960, Burgers 1969, Muchmore 1984, 
Iben \& MacDonald 1985) the multi-fluid equations

$$
{{dp_i} \over {dr}}-{{\rho _i} \over \rho }{{dp} \over {dr}}-n_iZ_ieE=
\sum\limits_j {K_{ij}}\left( {w_j-w_i} \right)\quad \quad \eqno\stepeq
$$

\noindent
where $p_i$, $\rho_i$, $n_i$, $Z_i$, $w_i$ are the partial pressure, mass 
density, number density, mean charge, and diffusion velocity for species $i$ 
and $E$ is the electric field induced by the gradients in the ion 
densities. The resistance coefficients are from Paquette at al (1986a).

In solving equations (1) for the diffusion velocities, care must be taken 
to avoid rounding errors due to the subtractions. This is particularly 
important for the diffusion velocities of trace elements. 
For $N$ species including electrons, equations (1) is a set of $N-1$ 
independent linear equations for $N+1$ unknown variables (the $w_{i}$'s 
and $E$). To complete the set of equations, we use the conditions for no 
net mass flow relative to the center of mass

$$
\sum\limits_i {A_i}n_iw_i=0\quad \quad \eqno\stepeq 
$$

\noindent
and charge neutrality

$$
\sum\limits_i {Z_i}n_i=0\quad \quad \eqno\stepeq 
$$
 
\noindent
In equations (2) and (3), the sums are over all species (ions, neutrals 
and electrons). We assume the electron to have zero mass. The electrical 
field can then be eliminated directly since equation (1) applied to 
electrons gives

$$
E=-{1 \over {en_e}}{{dp_e} \over {dr}}\quad \quad \eqno\stepeq
$$

\noindent
The electron pressure, $p_e$, and the electron number density, $n_e$, can 
now be eliminated since 

$$
p_e=p-\sum\limits_{i\ne e} {p_i}\quad \quad \eqno\stepeq
$$

\noindent
and

$$
n_e=\sum\limits_{i\ne e} {Z_i}n_i\quad \quad \eqno\stepeq
$$

\noindent
where the sums now do not include the electrons. Using the equation of 
hydrostatic balance, equation (1) can be transformed to

$$
\eqalignno{
{{dp_i} \over {dp}}-X_i
& +{{n_iZ_i} \over {\sum\limits_{j\ne e} {n_jZ_j}}}
\left[ {1- \sum\limits_{j\ne e} {{{dp_j} \over {dp}}}} \right]= \cr
& -\sum\limits_{j\ne e} {{{K_{ij}} \over {\rho g}}}\left( {w_j-w_i} \right)
\quad \quad 
& \stepeq \cr}
$$

\noindent
To reduce rounding errors, we reformulate equation (7) as

$$
\eqalignno{
\left( {{{dp_i} \over {dp}}-X_i} \right){{\sum\limits_{j\ne i,j\ne e} 
{{{X_jZ_j} \over {A_j}}}} \over {\sum\limits_{j\ne e} {{{X_jZ_j} \over 
{A_j}}}}}
& -{{X_iZ_i} \over {A_i}}{{\sum\limits_{j\ne i,j\ne e} 
{\left( {{{dp_j} \over {dp}}-X_j} \right)}} \over 
{\sum\limits_{j\ne e} {{{X_jZ_j} \over {A_j}}}}} = \cr
& -\sum\limits_{j\ne e} {{{K_{ij}} 
\over {\rho g}}}\left( {w_j-w_i} \right)\quad \quad 
& \stepeq \cr}
$$
 
\noindent
The equations for the changes in mass fractions due to element diffusion 
and convective mixing are

$$
{{dX_i} \over {dt}}+{{\partial F_i} \over {\partial m}}=0\quad \quad 
\eqno\stepeq
$$
 
\noindent
where the fluxes, $F_i$, are

$$
F_i=4\pi r^2\rho w_iX_i+\sigma _{con}{{\partial X_i} \over {\partial m}}
\quad \quad \eqno\stepeq
$$

\noindent
Convective mixing is treated as a diffusion process, with diffusion 
coefficient, $\sigma_{\rm con} = (4 \pi r^2 \rho l)^2 /\tau_{\rm con}$ 
where $l$ is the mixing length and $\tau_{\rm con}$ is the convective 
turnover time scale calculated from mixing length theory.

Equations (9) and (10) for each element are solved together with the 
equations of stellar structure by standard implicit finite difference 
techniques. The diffusion velocities, $w_i$, are first found by standard 
matrix methods (LU decomposition), from finite difference equations derived 
from equations (8). 

This multi-fluid formulation requires knowledge of the 
partial pressures, $p_i$. For non-interacting species or species interacting 
through short-range forces such as Van der Waals forces, calculation of 
$p_i$ is relatively straightforward. However for species interacting through 
long-range coulomb forces, it is not clear that the concept of partial 
pressures is meaningful. We circumvent this difficulty by adopting a `linear 
mixing law',

$$
p_i=n_i{{\sum\limits_{j\ne e} {p_j}} \over {\sum\limits_{j\ne e} {n_j}}}
\quad \quad \eqno\stepeq
$$

\begintable*{1} 
\caption{{\bf Table 1.} Properties of the initial models}
\halign{#\hfil           & \quad \hfil#\hfil\quad &
        \hfil#\hfil\quad &       \hfil#\hfil\quad & \hfil#\hfil\quad &
        \hfil#\hfil\quad &       \hfil#\hfil\quad & \hfil#\hfil\quad &
        \hfil#\hfil\quad &       \hfil#\hfil\quad & \hfil#\hfil      \cr
 Sequence & $M_{\rm core}$  & $X_{\rm C,core}$ & $X_{\rm O,core}$ & 
            $M_{\rm env}$   & $X_{\rm C,env}$  & $X_{\rm O,env}$  & 
            $R$             & $T_{\rm eff}$    & $\rho_{\rm c}$   & 
            $T_{\rm c}$     \cr
P6HE6 & $0.600$ & $0.50$ & $0.50$  & $4.93\times 10^{-7}$ &
        $0.01$  & $0.01$ & $1.13$  & $7.45$ & $2.78$  & $1.07$ \cr
P6HE4 & $0.600$ & $0.50$ & $0.50$  & $9.36\times 10^{-5}$ & 
        $0.01$  & $0.01$ & $1.16$  & $7.40$ & $2.78$  & $1.07$ \cr
P6HE3 & $0.600$ & $0.50$ & $0.50$  & $9.60\times 10^{-4}$ &
        $0.01$  & $0.01$ & $1.14$  & $7.45$ & $2.80$  & $1.07$ \cr
P6HE2 & $0.602$ & $0.50$ & $0.50$  & $8.29\times 10^{-3}$ &
        $0.01$  & $0.01$ & $1.13$  & $7.43$ & $2.97$ & $1.06$ \cr
P6LO3 & $0.600$ & $0.99$ & $0.01$  & $1.02\times 10^{-3}$ &
        $0.01$  & $0.01$ & $1.16$  & $7.49$ & $2.69$ & $1.03$ \cr
P6LO2 & $0.602$ & $0.99$ & $0.01$  & $8.29\times 10^{-3}$ &
        $0.01$  & $0.01$ & $1.15$  & $7.47$ & $2.86$ & $1.07$ \cr
P6PG6 & $0.600$ & $0.50$ & $0.50$  & $4.41\times 10^{-6}$ & 
        $0.50$  & $0.17$ & $1.88$  & $10.97$ & $1.40$ & $2.02$ \cr
1HE4  & $1.00$  & $0.50$ & $0.50$  & $1.01\times 10^{-4}$ & 
        $0.01$  & $0.01$ & $0.590$ & $7.68$ & $30.7$ & $1.08$ \cr
1HE3  & $1.00$  & $0.50$ & $0.50$  & $1.02\times 10^{-3}$ & 
        $0.01$  & $0.01$ & $0.590$ & $7.73$ & $30.9$ & $1.10$ \cr
}
\tabletext{Units are $M_{\odot}$ for masses, $10^{9}$ cm for $R$, 
$10^4$ K for $T_{\rm eff}$, $10^6$ g~cm$^{-3}$ for $\rho_{\rm c}$, and 
$10^8$ K for $T_{\rm c}$.} 
\endtable

\section{Evolution of the abundance profiles}

Figure 1 shows how the carbon mass fraction changes with $T_{\rm eff}$ for 
the P6HE3 sequence. Also shown are the location of the photosphere and the 
boundaries of the convection zones. This figure can be compared to figure 1 
of P86. The most important difference is in $M_{\rm cmax}$, the maximum 
depth of the inner boundary of the convection zone. For 
$M_{*} = 0.6 M_{\odot}$, and $M_{\rm env} = 1.9~10^{-4} M_{\odot}$, 
P86 find $M_{\rm cmax} = 3.8~10^{-7} M_{\odot}$, which occurs when 
$T_{\rm eff}$ is near $10^{4}$ K. For the P6HE4 and P6HE3 sequences, we 
find $M_{\rm cmax} = 1.8~10^{-6}$ and $1.7~10^{-5} M_{\odot}$, respectively. 
We attribute most of the order of magnitude difference in $M_{\rm cmax}$ to 
differences in radiative opacity. Although the depth of the convection zone 
is sensitive to convective efficiency for relatively high effective 
temperatures, Fontaine, Tassoul \& Wesemael (1984) have shown that at 
temperatures characteristic of DQs the dependence is significantly less. To 
illustrate this, we have calculated cooling sequences without 
element diffusion for a $0.61 M_{\odot}$ white dwarf with $M_{\rm env} = 
10^{-2} M_{\odot}$ and mixing length ratios, $\alpha$ = 1.0, 1.5 and 2.0. 
In figure 2, we show the location of the convection zone inner boundary for 
pure helium envelopes and envelopes that are a mixture of helium and carbon 
with $X_{\rm C} = 10^{-3}$. To test the dependence of $M_{\rm cmax}$ on 
opacity, we have also calculated cooling sequences using the Los 
Alamos opacities (Huebner et al. 1977). In figure 3, we show the location 
of the convection zone inner boundary for the helium and carbon mixture for 
the OPAL and Los Alamos opacities. It is clear that the newer opacities 
result in deeper convection zones and that the depth of the convection zone 
is insensitive to $\alpha$ for $T_{\rm eff} < 14,000$ K. We also see that 
the presence of carbon in small amounts reduces the depth of the convection 
zone, and hence the convection zone in the P6HE4 sequence is smaller than 
that of the P6HE3 sequence.

\beginfigure{1}
\epsfxsize=9cm
\epsffile{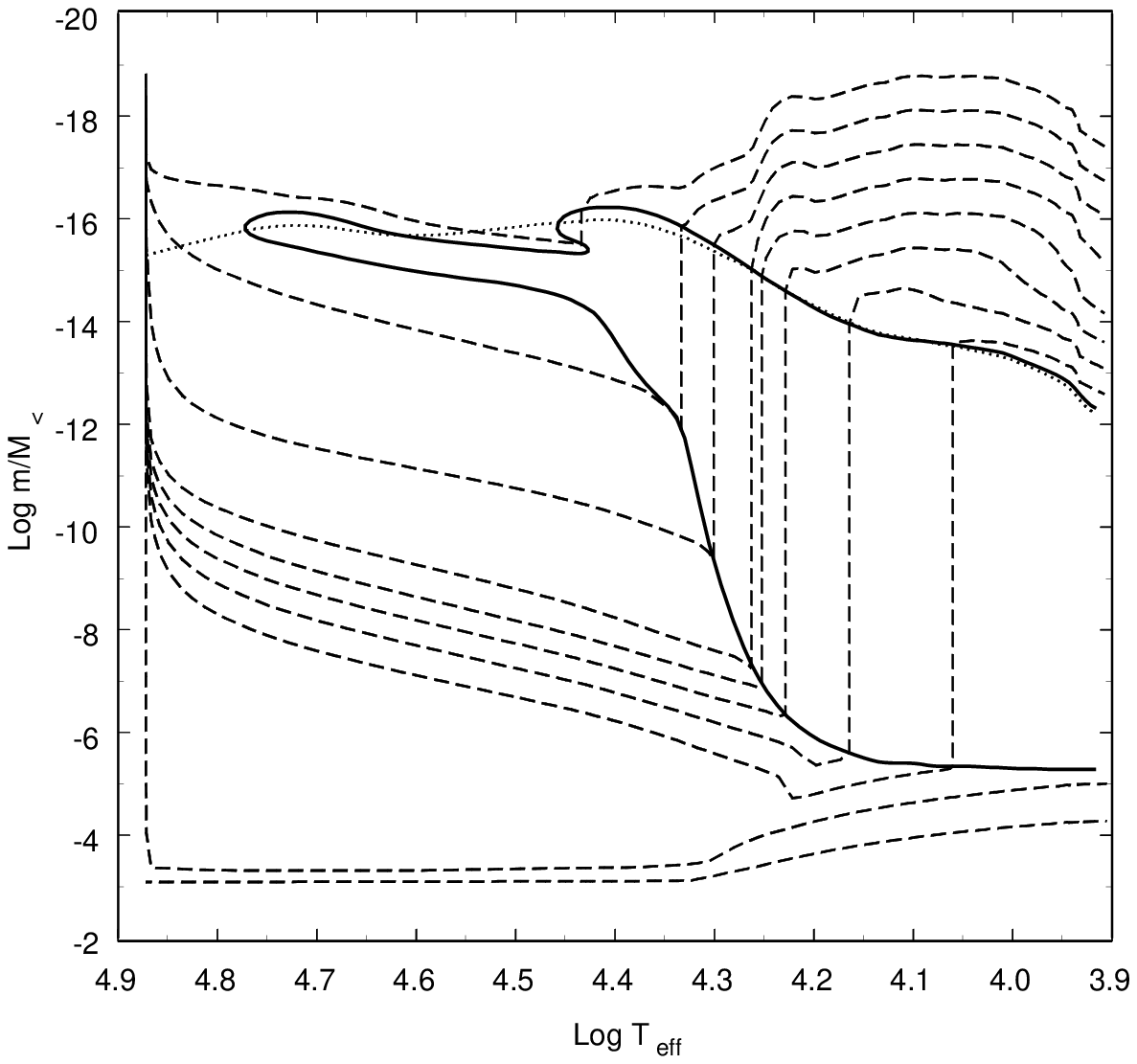}
\caption{{\bf Figure 1.} Convection zone boundaries (solid line), location
     of the photosphere (dotted line) and contours of carbon mass fraction
     (dashed lines) for the P6HE3 sequence. The contour levels are, from top
     to bottom, log $X_{\rm C12} = -10, -9, ..., -1$.}
\endfigure

\beginfigure{2}
\epsfxsize=9cm
\epsffile{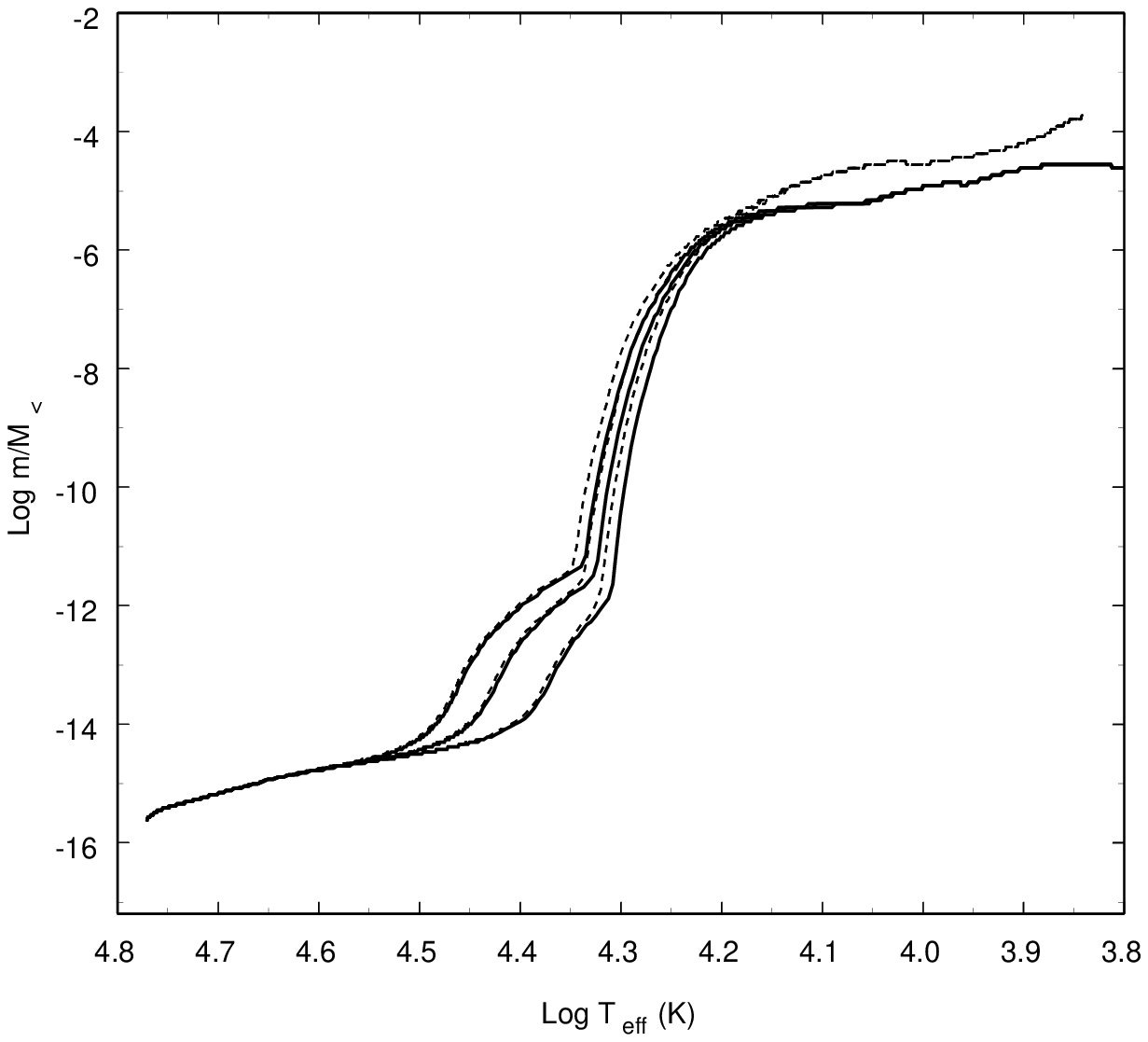}
\caption{{\bf Figure 2.} Location of the inner boundary of the convection
     zone for a $0.61 M_{\odot}$ white dwarf with $M_{\rm env} = 10^{-2}
     M_{\odot}$ and mixing length ratios, $\alpha$ = 1.0, 1.5 and 2.0 (from
     bottom to top). The solid lines are for $X_{\rm C} = 10^{-3}$ and the
     dashed lines are for $X_{\rm C} = 0$.}
\endfigure

\beginfigure{3}
\epsfxsize=9cm
\epsffile{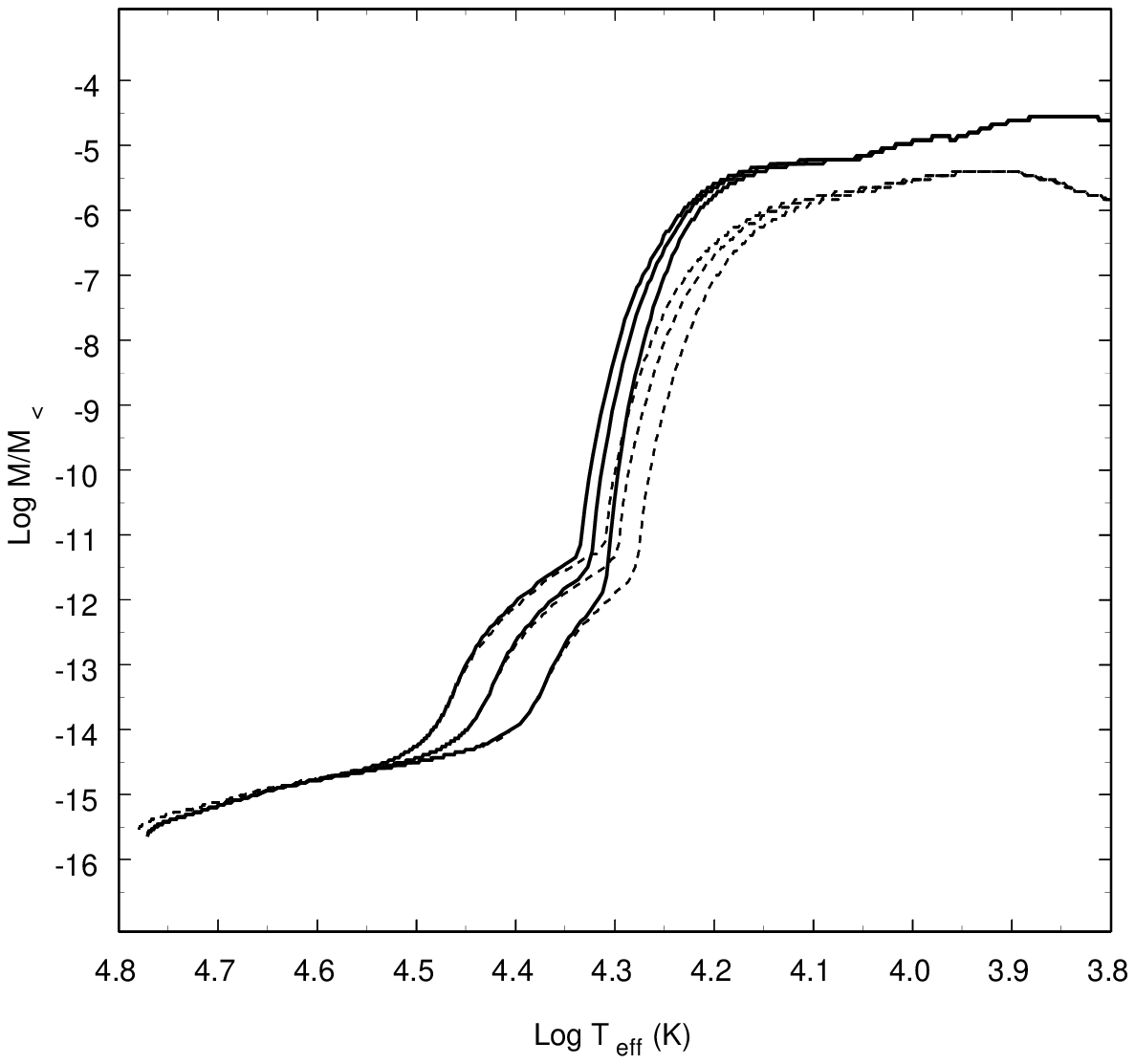}
\caption{{\bf Figure 3.} Location of the inner boundary of the convection
     zone for a $0.61 M_{\odot}$ white dwarf with $M_{\rm env} = 10^{-2}
     M_{\odot}$ and mixing length ratios, $\alpha$ = 1.0, 1.5 and 2.0 (from
     bottom to top). The solid (dashed) lines are for when OPAL (Los Alamos)
     opacities are used.}
\endfigure

\subsection{Comparison with observations and prior evolution}

In figure 4, we show the ratio of the number densities of carbon and helium 
in the convection zone as a function of $T_{\rm eff}$ for the models with 
$M_{\rm core} = 0.6 M_{\odot}$. Also shown are the observed photospheric 
ratios taken from Weidemann \& Koester (1995). When possible, error bars 
are from the individual papers referenced in Weidemann \& Koester (1995). 
Excluding the stars for which there are only upper limits, the mean of 
the observed C to He ratios is log $n$(C)/$n$(He) = $-5.35$ and the 
corresponding mean $T_{\rm eff}$ is 9,300 K. For our models with $M_{\rm 
core} = 0.6 M_{\odot}$, these values are obtained if $M_{\rm env} = 
1.1~10^{-2} M_{\odot}$. 

There are many possible ways that helium atmosphere white dwarfs can form. 
They may be descendants of asymptotic giant branch (AGB) stars or sdOs. They 
may also be the result of mergers in double degenerate systems. Because the 
evolutionary state of sdO stars is poorly understood and the physics of 
mergers is very complex, we assume here that DQs have AGB predecessors. 

The possible modes of how AGB stars evolve to white dwarfs have been 
described in detail by Iben (1984). In contrast to the case for DA white 
dwarfs, there have been too few detailed calculations of the evolution to 
helium atmosphere white dwarfs for us to be precise about the mass of helium 
in these objects. However due to the relationship between the white dwarf 
and its AGB predecessor, we find it convenient to parameterize $M_{\rm env}$ 
in terms of $M_{\rm He}^{\rm init}$, the mass of the helium layer just prior 
to ignition of a helium shell flash. This is given by (Iben \& Tutukov 1996)

$$
\log M_{\rm He}^{\rm init}=-1.835+1.73M_{\rm core}-2.67M_{\rm core}^2
\quad \quad \eqno\stepeq
$$

\noindent
for $M_{\rm core} \ge 0.55 M_{\odot}$. This is in good agreement with the 
results for the final helium layer mass from unpublished calculations by one 
of us (JM) of the evolution of population I helium stars of masses between 
0.5 and 0.8 $M_{\odot}$ to the white dwarf stage. For less massive cores, we 
find that this equation underestimates $M_{\rm He}^{\rm init}$.

Whether the final white dwarf has a hydrogen-deficient envelope or not 
depends on the phase of the thermal pulse cycle at which the star leaves 
the AGB. If this occurs during helium shell burning, winds can remove all 
remaining hydrogen. Thus we might expect $M_{\rm env}$ to be close to
$M_{\rm He}^{\rm end}$, the mass of the helium layer at the end of 
quiescent helium burning, in a star that is near the tip of the AGB 
(D'Antona \& Mazzitelli 1991). This is about 30\% of $M_{\rm He}^{\rm init}$ 
(Vassiliadis \& Wood 1993; Iben \& Tutukov 1996). D'Antona \& Mazzitelli 
(1991) estimate 
that $M_{\rm env}$ will be about 30\% less than $M_{\rm He}^{\rm end}$ 
because carbon and other heavy elements will sink out of the helium buffer 
layer as the star evolves to a white dwarf. However, Iben \& Tutukov (1984) 
find for $M_{*} = 0.6 M_{\odot}$, that $M_{\rm env} = 0.016 M_{\odot}$, 
which is about 0.9 $M_{\rm He}^{\rm init}$. Thus $M_{\rm env}/
M_{\rm He}^{\rm init}$ is of order 0.2 to 1.0 so that for $M_{\rm core} = 
0.6 
M_{\odot}$, we estimate $M_{\rm env}$ is between $3.5~10^{-3}$ and 
$1.7~10^{-2} M_{\odot}$, which is in reasonably agreement with 
the mean value derived above for the DQ white dwarfs. Furthermore this 
range of $M_{\rm env}$ corresponds to a range in log $n$(C)/$n$(He) at 9,300 
K 
of $-4.2$ to $-5.8$. Although many of the DQs lie within these limits, 
clearly some of the data points fall outside this range. The DQ white dwarf 
with the largest C to He ratio, G35-26, has log $n$(C)/$n$(He) = $-1.5$ and 
$T_{\rm eff} \approx 12,500$ K. If $M_{\rm core} = 0.6 M_{\odot}$, this 
would require $M_{\rm env} = 2~10^{-5} M_{\odot}$. At the other extreme, 
there are two DZ, two DB and 1 DBA white dwarfs with upper limits to 
log $n$(C)/$n$(He) of less than $-6.5$.

We suggest that DQs with very high C to He ratios are significantly more 
massive than the typical white dwarf. Due to the higher gravity these have 
smaller scale heights in the outer layers, which leads to less massive 
helium 
envelopes. Figure 5 is the same as figure 4 except that the core mass is 
1.0 $M_{\odot}$. The solid lines are for envelope masses of $10^{-4}$ and 
$10^{-3} M_{\odot}$. The range in $M_{\rm env}$ consistent with prior 
evolution is $5.0~10^{-4} < M_{\rm env} < 1.7~10^{-3} M_{\odot}$. 
Extrapolating our results at 0.6 and 1.0 $M_{\odot}$, we estimate that 
the minimum core mass consistent with the observed C to He ratio 
for G35-26 is about 1.3 $M_{\odot}$, which is consistent with the high 
gravity found by Thejll et al. (1990) for this star from analysis of 
spectra. 
White dwarfs of  mass greater than 1.2 $M_{\odot}$, however, are 
thought to evolve from stars of initial mass between 7.5 and 10 $M_{\odot}$ 
(with an uncertainty of about 0.5 $M_{\odot}$ in both limits) which 
experience quiescent core carbon burning and so develop ONe cores 
(Dom\'\i nguez, Tornamb\'e \& Isern, 1993; Ritossa, Garc\'\i a-Berro \& Iben 
1996). Hence we can only conclude that the white dwarf in G35-26 is more 
massive than 1 $M_{\odot}$. 

Similarly the lowest C to He ratios can be explained by allowing the white 
dwarf to be less massive than the canonical 0.6 $M_{\odot}$. The maximum 
$M_{\rm env}$ for a 0.5 $M_{\odot}$ white dwarf is 0.04 $M_{\odot}$. 
With this much helium, we estimate that $n$(C)/$n$(He)$\sim 10^{-7}$, which 
is 
about the lowest detected carbon abundance.

In contrast to P86, we do not find that the carbon abundance starts to 
drop near $T_{\rm eff} = 7,000$ K. This difference is due to carbon being 
fully ionized below the convection zone in the P6HE4, P6HE3 and P6HE2 
sequences, whereas P86 find partial recombination. In their 
calculations, P86 used the Fontaine, Graboske and Van Horn (1977) 
equation of state. Our equation of state is similar in that it is also 
based on free energy minimization and is thermodynamically consistent. An 
important difference that might be the cause of the difference in 
ionization structure is in the treatment of Stark ionization i.e. the 
dissolution of bound states due to the ambient electric field from the 
presence of free charges. We use the occupancy probability formalism of 
Hummer and Mihalas (1988). Fontaine et al (1977) divide the temperature-
density plane into three regions. In a low density regime, they truncate 
the partition function by summing over the finite number of bound state 
energy levels calculated from the static screened coulomb potential. In a 
high density regime they assume complete ionization. In the intermediate 
region they interpolate between the low and high density regimes. In 
table 2, we give some relevant properties of the convection zones for the 
coolest model of each of the HE sequences. In each case, conditions are 
such that they lie in the high density fully ionized regime of the 
Fontaine et al (1977) equation of state. We have already noted that P86 
find shallower convection zones. At a mass depth of $4.0~10^{-7} 
M_{\odot}$, corresponding to the depth of the base of the convection zone 
in the model of P86 with $M_{\rm env} = 1.9~10^{-4} M_{\odot}$, the 
temperature is $1.7~10^6$ K and the density is $7.7~10^1$ g cm$^{-3}$ for 
the P6HE4 model, which again places this point in the high temperature 
fully ionized region of the Fontaine et al (1977) equation of state. 
Hence we conclude that the difference in ionization structure in the 
layers neighboring the base of the convection zone must be due to higher 
temperatures in our models than those of P86.

\begintable*{2} 
\caption{{\bf Table 2.} Properties of the convective zones for the 
coolest model of each HE sequence}
\halign{#\hfil           & \quad \hfil#\hfil\quad &
        \hfil#\hfil\quad &       \hfil#\hfil\quad & \hfil#\hfil\quad &
        \hfil#\hfil\quad &       \hfil#\hfil\quad & \hfil#\hfil      \cr
 Sequence & $T_{\rm eff}$   & $M_{\rm conv}$   & $T_{\rm conv}$   & 
            $\rho_{\rm conv}$     \cr
P6HE6 & $9,930$ & $1.973\times 10^{-7}$ & $3.08$  & $33.7$ \cr
P6HE4 & $7,640$ & $6.950\times 10^{-7}$ & $2.26$  & $117$  \cr
P6HE3 & $6,830$ & $7.015\times 10^{-6}$ & $1.07$  & $740$  \cr
P6HE2 & $6,960$ & $7.621\times 10^{-5}$ & $0.857$ & $3610$ \cr
}
\tabletext{Units are $M_{\odot}$ for masses, 
K for $T_{\rm eff}$, $10^6$ K for $T_{\rm conv}$, and $10^2$ g~cm$^{-3}$ 
for $\rho_{\rm conv}$.} 
\endtable

\beginfigure{4}
\epsfxsize=9cm
\epsffile{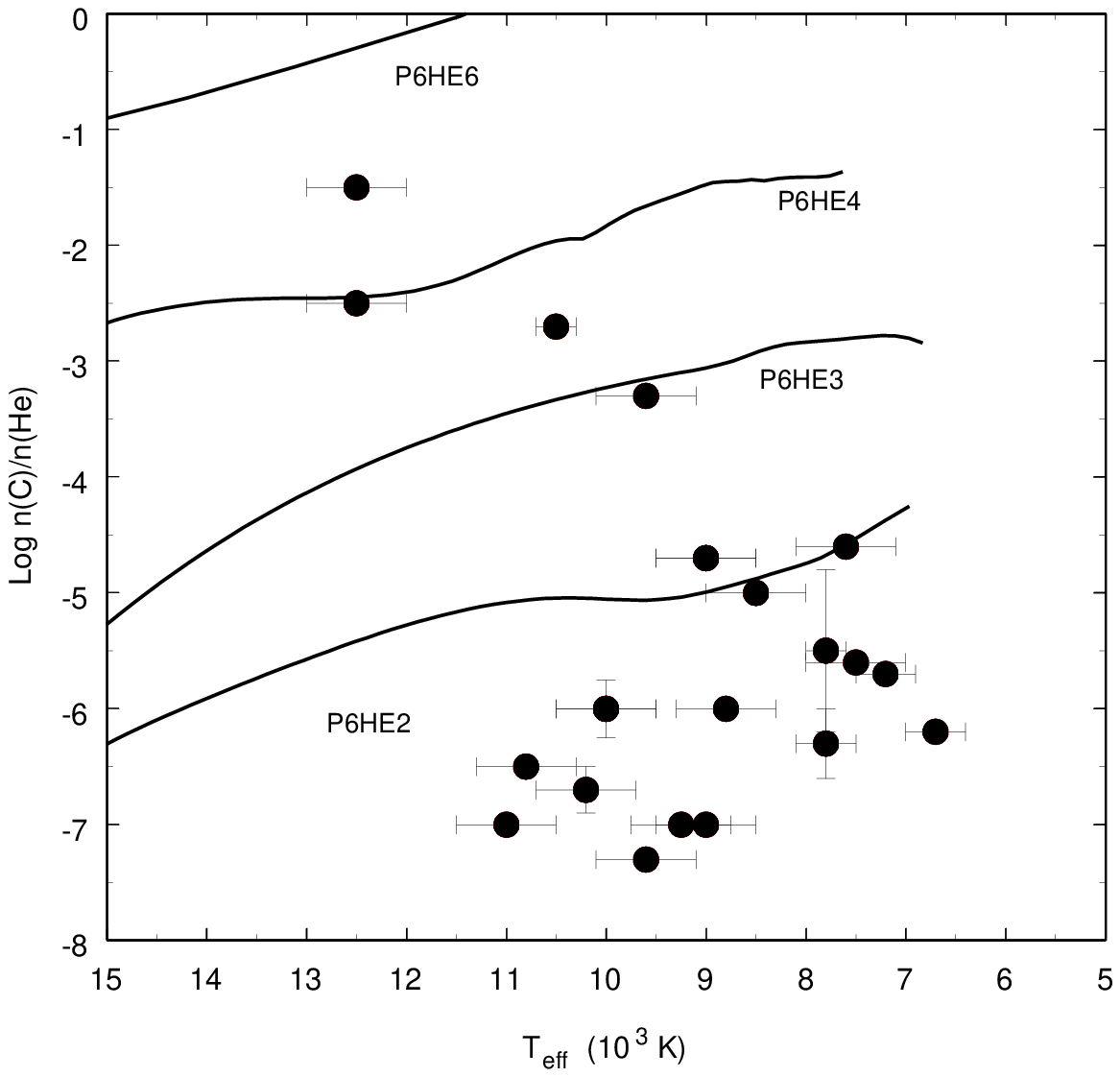}
\caption{{\bf Figure 4.} The solid lines are the values of log 
$n$(C)/$n$(He)
     for the P6HE6, P6HE4, P6HE3 and P6HE2 sequences. Also
     shown with error bars are the observed ratios.}
\endfigure

\beginfigure{5}
\epsfxsize=9cm
\epsffile{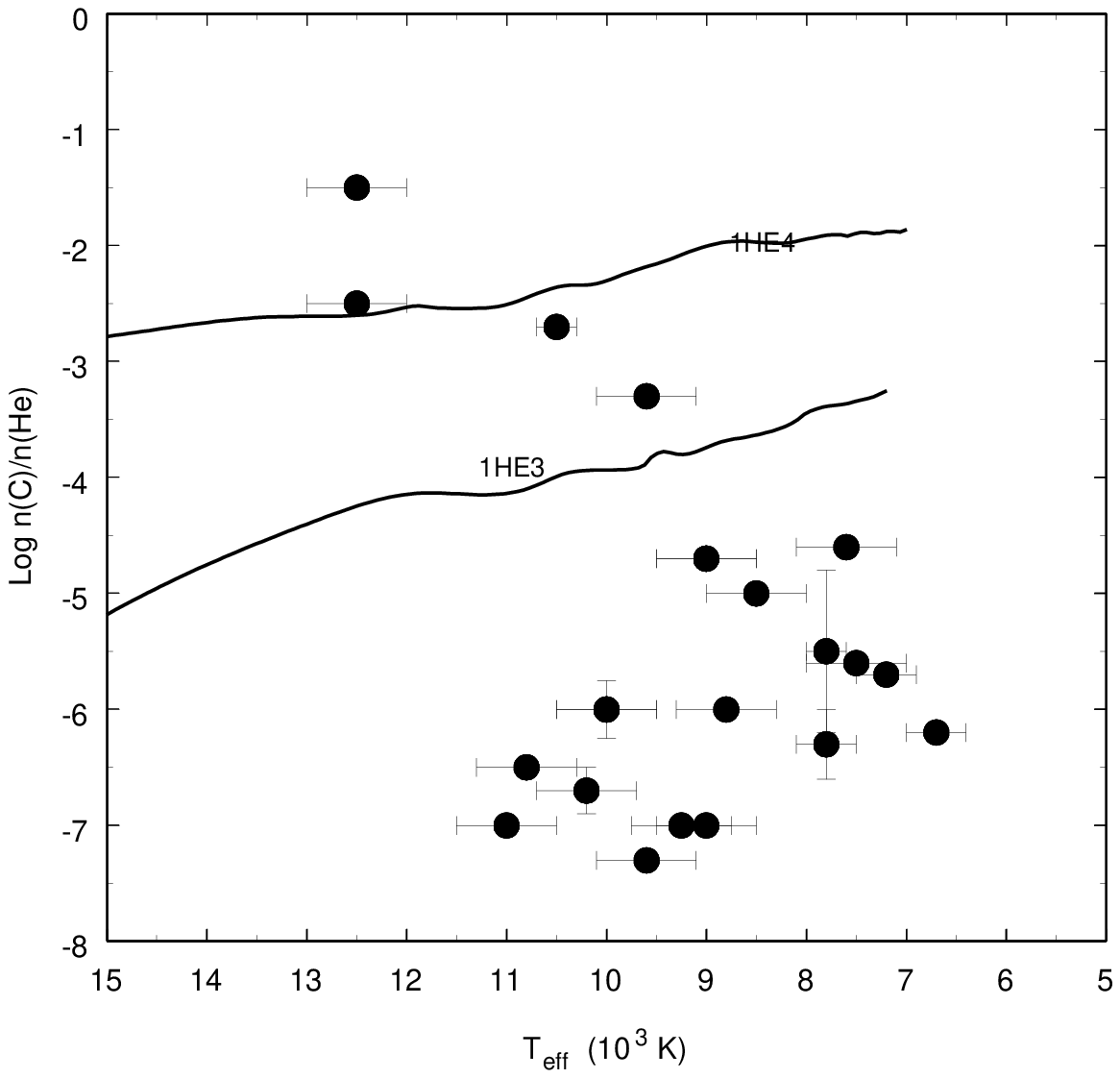}
\caption{{\bf Figure 5.} As figure 4 but for the 1HE4 and 1HE3
     sequences.}
\endfigure

\subsection{The role of oxygen}

The mass fraction of oxygen in the white dwarf core has a strong dependence 
on the rate of the $^{12}$C$(\alpha,\gamma)^{16}$O reaction. Although 
recent experimental progress has reduced the error in this reaction 
rate to about 35\%, this uncertainty can still be significant for 
nucleosynthesis studies and models of type I supernovae (Arnett 1996). 
Hence a measurement of the oxygen abundance in the atmosphere of a DQ white 
dwarf may provide useful information about this reaction rate. 

Salaris et al. (1997) have recently calculated the evolution of population 
I stars of mass 3.2 and 7.0 $M_{\odot}$, using the best current estimate 
for the $^{12}$C$(\alpha,\gamma)^{16}$O reaction rate (Woosley, Timmes \& 
Weaver 1993; Thielemann, Nomoto \& Hashimoto 1996; Arnett 1996). These stars 
develop CO cores of mass 0.6 and 1.0 $M_{\odot}$. The central mass fractions 
are $X_{\rm C} = 0.233, X_{\rm O} = 0.739$ and $X_{\rm C} = 0.316, 
X_{\rm O} = 0.655$ respectively. In our unpublished calculations of the 
evolution of population I helium stars, using the Arnett (1996) reaction 
rate, 
we find slightly higher central oxygen abundances. In the context of dredge-
up 
of oxygen diffusing from the core a more relevant quantity is the O mass 
fraction just below the extinct helium burning shell. The Salaris et al. 
(1997) 
models give $X_{\rm O} \approx 0.02$ at the center of the helium shell. For 
the population I helium stars, we find $X_{\rm O} \approx 0.05$ at the same 
location. 

To get some idea of how the abundance of oxygen in the core affects the 
amount of oxygen dredged up to the surface, we have made two additional 
evolutionary sequences, P6LO3 and P6LO2 for 0.6 $M_{\odot}$ cores of 
composition $X_{\rm C} = 0.99, X_{\rm O} = 0.01$. Although the abundance 
profiles in our white dwarf models differ in detail from those of the 
evolutionary calculations, the two values of core oxygen abundance 
considered 
here roughly bracket the possible range of oxygen abundance in the region 
below the helium shell. In figure 6, we show the O to He ratio as a 
function of $T_{\rm eff}$ for these two sequences and the P6HE2 and P6HE3 
sequences. In the sequences with the thicker helium layer very little oxygen 
is dredged to the surface, independent of the amount of oxygen in the core. 
For both of the thinner helium envelopes, oxygen is expected to be present 
in detectable amounts (based on a comparison of the oscillator strengths 
of the O I multiplets at 130.5 and 777.4 nm with the detected C I lines). 
Also the maximum photospheric oxygen abundance depends on the core oxygen 
abundance. In the P6HE3 sequence with $X_{\rm O,core}$ = 0.5, the maximum 
photospheric abundance is log $n$(O)/$n$(He) = $-4.0$ which occurs at 
$T_{\rm eff} =  8.0~10^{3}$ K and in the P6LO3 sequence with 
$X_{\rm O,core}$ = 0.01, the maximum photospheric abundance is 
log $n$(O)/$n$(He) = $-5.3$ which occurs at $T_{\rm eff} = 8.3~10^{3}$ K. 
Hence measurement of the photospheric oxygen abundance in a DQ has the 
potential of providing useful information about the interior composition of 
the white dwarf.

\beginfigure{6}
\epsfxsize=9cm
\epsffile{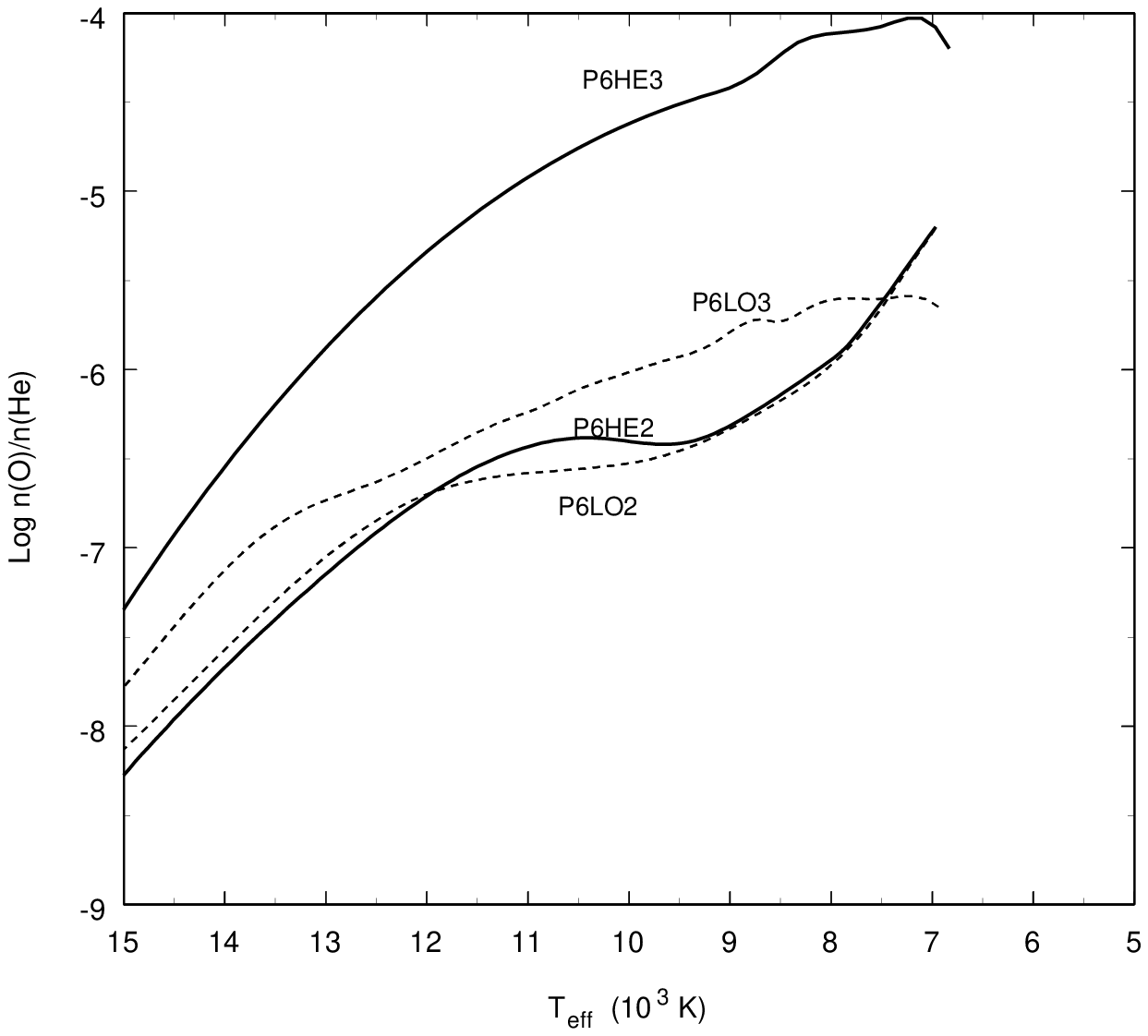}
\caption{{\bf Figure 6.} The values of log $n$(O)/$n$(He) for the P6HE3 and
     P6HE2 (solid lines) and the P6LO3 and P6LO2 (dashed lines) sequences.}
\endfigure

\subsection{The DBV GD~358}

UV spectroscopy of the DBV GD~358 by Provencal et al. (1996) has confirmed 
the detection of He II 164.0 nm and C II 133.5 nm, first reported by 
Sion et al. (1988). Thus GD~358 can be considered to be the hottest DQ star, 
with $T_{\rm eff} \approx  27,000$ K and log $n$(C)/$n$(He) = $-5.65 \pm 
0.10$ 
(Provencal et al. 1996). At this temperature, the star is sufficiently old 
that heavy elements will have sunk deep beneath the photosphere yet too hot 
for a deep convection zone to develop and dredge-up carbon diffusing out of 
the core, unless the helium layer is very thin.

Dehner \& Kawaler (1995) have investigated the possibility that GD~358 is 
a descendant of PG1159. They calculate the evolution of the abundance 
profiles in a star that has an initial envelope of mass $3~10^{-3} 
M_{\odot}$ 
and composition 30\% He, 35\% C and 35\% O by mass. When the star has cooled 
to the temperature range of the DBV's, they find that the star has a layer 
of pure helium of mass $10^{-5.5} M_{\odot}$, in good agreement with the 
asteroseismological analysis of the WET observations by Bradley \& Winget 
(1994). Our P6PG6 sequence begins with a helium mass comparable to that 
found by asteroseismology for GD~ 358. When $T_{\rm eff} = 27,000$ K, 
the center of the transition from He to C is at a mass depth of 
$10^{-6.7} M_{\odot}$, i.e. an order of magnitude less than found from 
asteroseismology, yet the photospheric $n$(C)/$n$(He) is $\sim 10^{-20}$, 
which is many orders of magnitude smaller than observed. To get a detectable 
C to He ratio at 27,000 K, the helium layer mass cannot be too much larger 
than the convection zone, which at this temperature has its base 
at mass depth of $10^{-14} M_{\odot}$.  Hence scenarios in which GD~358 is 
a descendant of the PG1159 star suffer from a conflict between the 
requirement that the helium layer mass be consistent with asteroseismology 
and the requirement that it be very small for C to be present in the 
photosphere at the observed abundance.

Other possible sources for the photospheric carbon are: 1) that it is 
radiatively supported, 2) accretion from the ISM and 3) that it is left 
over from a merger of two low mass white dwarfs in a double degenerate 
binary.

Detailed calculations of radiative levitation in helium dominated 
atmospheres 
have recently been presented by Chayer, Fontaine \& Wesemael (1995), who 
find that, for $T_{\rm eff} < 4~10^{4}$ K, the radiatively levitated 
abundance of carbon is significantly less than $n$(C)/$n$(He) = $10^{-6}$. 
Hence radiative levitation appears to be insufficient to support carbon at 
the levels observed in GD~358.

Clearly, if accretion is invoked as the mechanism for the presence of 
photospheric carbon, then an explanation of why hydrogen has a low 
photospheric abundance of $n$(H)/$n$(He) $\leq 3~10^{-5}$ (Provencal et al. 
1996) is needed.  Mechanisms that prevent or inhibit accretion have been 
reviewed by Alcock \& Illarionov (1980). However most of these mechanisms 
work 
equally well for carbon as for hydrogen. One mechanism that permits 
differential accretion is that the white dwarf  has sufficiently high 
rotation 
rate and magnetic field that the `propeller' mechanism (Illarionov \& 
Sunyaev 
1975) reduces the accretion rate of ionized hydrogen below the Bondi-Hoyle 
rate. A significant fraction of the carbon in the interstellar medium (ISM) 
can 
be in the form of grains, which have a significantly lower charge to mass 
ratio than protons so that they follow independent particle trajectories 
while outside the white dwarf's magnetosphere (Alcock \& Illarionov 1980). 
If they cross into the magnetosphere and are then evaporated by the 
radiation 
from the white dwarf, they provide a source of ionized carbon and other 
species that can then follow magnetic field lines to the white dwarf's 
surface. The requirements for an effective propeller are discussed by 
Alcock \& Illarionov (1980) and Stella, White \& Rosner (1986). Accretion 
of ionized hydrogen is centrifugally inhibited if the co-rotation radius, 
$r_{\rm c}$, is less than the magnetospheric radius, $r_{\rm m}$. By 
studying  multiplet splitting in their WET observations of the oscillations 
of GD~358, Winget et al. (1994) find evidence for a magnetic field of 
average 
strength 1300$\pm$ 300 G and for differential rotation with the outer 
envelope rotating with a period of 0.89 d. With these parameters, 
$r_{\rm c} \approx 5~10^{9}$ cm. To determine $r_{\rm m}$, we equate the 
magnetic energy density to the kinetic energy density on the symmetry axis 
of the (upstream) accretion flow. Assuming a dipole magnetic field of 
strength $B_{*}$,

$$
\eqalignno{
r_m= 
& 8\;10^{10}\left( {{{B_*} \over {10^3{\rm G}}}} \right)^{4/9}\left( {{{R_*} 
\over {10^9{\rm cm}}}} \right)^{4/3}\left( {{{v_\infty } 
\over {50{\rm km\,\rm s^{-1}}}}} \right)^{2/9} \cr
& n_\infty ^{-2/9}\left( {{{M_*} \over {M_\odot}}} \right)^{-1/3} \, {\rm 
cm}
\quad \quad 
& \stepeq \cr}
$$

\noindent
where $v_\infty$ and $n_\infty$ are the velocity of the ISM relative to star 
and the ISM number density respectively. For GD~358, Provencal et al. (1996) 
find $v_\infty \approx 40$ km s$^{-1}$ and hence $r_{\rm m} > r_{\rm c}$ 
if $n_\infty < 3~10^{5}$  cm$^{-3}$, which is the case for much of the 
volume of the ISM. Hence the propeller mechanism is capable of inhibiting 
accretion of ionized material  by GD~358 and this may be the reason why the 
photospheric abundance of hydrogen is low. However, for carbon to be 
accreted at a greater rate than hydrogen, we also require that the 
characteristic distance at which incoming grains evaporate, $r_{\rm e}$, 
be less than $r_{\rm m}$. The temperature of dust grains of characteristic 
size 0.3 $\mu$m heated by the radiative flux from a star of distance $r$ 
and luminosity $L$ is (Osterbrock 1989)
 
$$
T_d\approx 10^3\left( {{L \over {0.1L_\odot}}} \right)^{1/5}\left( {{r \over 
{10^{13}{\rm cm}}}} \right)^{-2/5}{\rm K}\quad \quad \eqno\stepeq
$$ 

\noindent
For refractory grains which evaporate at $\approx 10^{3}$ K, 
$r_{\rm e} \approx 10^{13}$ cm which is much greater than $r_{\rm m}$ for 
any values of $n$ relevant to the ISM. Hence it is unlikely that the 
propeller mechanism can inhibit accretion of hydrogen without doing the 
same for carbon. From our P6PG6 sequence, we estimate that the time scale 
for settling of carbon out of the convection zone is about 0.1 yr. Hence, 
in a steady state, carbon has to be accreted at a rate of $4~10^{-19} 
M_{\odot}$ yr$^{-1}$ to maintain the observed photospheric abundance. For 
fluid dynamical accretion, and cosmic abundances this requires an ISM 
density 
of $n_{\rm H}$ = 5 cm$^{-3}$, which is higher than in the local ISM.

The merging of two white dwarfs in a double degenerate scenario is a 
complex process that is poorly understood. However it might reasonably 
be expected that the result of the merger is a luminous object that has 
much evolution to go through before reaching properties similar to those 
of GD~358. Hence given the short time scale for the evolution of the carbon 
distribution, it would be a remarkable coincidence that we find the star 
just at the time it looks like GD~358. 

Hence none of these mechanisms can satisfactorily explain the observed 
photospheric abundances in a way that is consistent with the data from 
asteroseismology.

\section{Conclusions and discussion}

Our main conclusion is that use of up-to-date opacities in modeling the 
evolution of DQ white dwarfs removes the discrepancy between estimates of 
the helium layer masses from the observed carbon abundances and the 
predictions of stellar evolution theory. In particular the bulk of the DQ 
white dwarfs can be interpreted as having mass $0.6 M_{\odot}$ and helium 
layer masses between $10^{-3}$ and $10^{-2} M_{\odot}$.

We suggest that DQ white dwarfs with exceptionally high atmospheric C 
abundances, such as G35-26, are of higher than average mass and those 
with no C or very low C have lower than average mass.

We also predict that oxygen will be present in DQ white dwarf atmospheres 
in detectable amounts if the helium layer mass is near the lower limit 
compatible with stellar evolution theory. Determination of the oxygen 
abundance has the potential of providing information on the profile of 
oxygen in the core and hence on the important 
$^{12}$C$(\alpha,\gamma)^{16}$O reaction rate. 

In contrast to P86, we do not find a drop in photospheric carbon 
abundance near $T_{\rm eff}$ = 7,000 K. Unfortunately, we are not able to 
continue our calculations below this temperature due to lack of opacity 
tables for low temperatures. Furthermore, for $T_{eff}$ below about 
10,000 K, uncertainties in the equation of state are becoming 
increasingly important. In particular, the equation of state is becoming 
sensitive to the treatment of Stark ionization. Hence we do not rule out 
the possibility that partial recombination of carbon occurs below the 
convection zone leading to a drop in photospheric carbon abundance as 
found by P86.

\section*{Acknowledgments}

JM thanks Judi Provencal for useful discussions and the Human Capital and
mobility programme (CHGE-CT92-0009) "Access to Supercomputing Facilities",
established between the EC and the CESCA/CEPBA. MH thanks the CIRIT for a
grant to visit the 
University of Delaware, as well as the Department of Physics \& Astronomy 
of the University of Delaware for hospitality during her stay.
We also thank partial support from the DGICYT Projects PB94-0111 and 
PB94-0827-C02-02.

\bye